\documentstyle[seceq,preprint]{jpsj}

\def\gsim{\,$\raise0.3ex\hbox{$>$}\llap{\lower0.8ex\hbox{$\sim$}}$\,}
\def\lsim{\,$\raise0.3ex\hbox{$<$}\llap{\lower0.8ex\hbox{$\sim$}}$\,}

\def\runtitle{Ground-State Phase Diagram of Frustrated Antiferromagnetic 
$S = 1$ Chain ...}
\def\runauthor{Toshiya {\sc Hikihara}}

\title{Ground-State Phase Diagram of Frustrated Antiferromagnetic 
$S = 1$ Chain with Uniaxial Single-Ion-Type Anisotropy}

\author
{Toshiya {\sc Hikihara}\footnote{E-mail address: 
HIKIHARA.Toshiya@nims.go.jp}}

\inst
{National Institute for Materials Science, Tsukuba, Ibaragi 305-0047}

\recdate{\today}

\abst
{The ground-state phase diagram of a frustrated $S = 1$ Heisenberg spin chain 
with uniaxial single-ion-type anisotropy is studied using 
the infinite-system density-matrix renormalization-group method.
Particular attention is paid to ^^ ^^ chiral" phases, 
in which the long-range order parameter is a vector chirality.
It is found that, 
in addition to the Haldane, large-D (LD), and double-Haldane phases,
the phase diagram includes three different chiral phases, i.e.,
the gapless chiral, chiral-Haldane, and chiral-LD phases.
These chiral phases are distinguished from each other 
by the energy gap and the string long-range order.
}

\kword
{frustration, spin-1 chain, ground-state phase diagram, 
 uniaxial single-ion-type anisotropy, chiral ordering, 
 density-matrix renormalization-group method}

\begin{document}
\sloppy
\maketitle

\section{Introduction}

The effect of geometrical frustration in quantum spin systems 
has attracted considerable attention for many years.
Frustration generally suppresses the tendency towards 
the classical N{\`e}el ordering and 
leads to a wide variety of exotic phenomena such as 
a spontaneous breaking of the translational symmetry 
in a one-dimensional (1D) system and 
spin-liquid states in 2D and 3D systems.
Among them, the possibility of a novel ^^ ^^ chiral" phase in 
1D frustrated quantum spin systems 
has been studied intensively in recent years.
~\cite{Ner,Leche,Kole,KKH,HKK1,HKK2,Nishi}
This chiral phase is characterized by 
the absence of the helical long-range order (LRO),
\begin{equation}
{\mib m}(q) = \frac{1}{LS} \sum_l {\mib S}_l {\rm e}^{{\rm i}ql} = 0 
~~~~({\rm for}~\forall~q), \label{eq:Ohel}
\end{equation}
and the non-zero value of the $z$-component of 
the total vector chirality,~\cite{scalchi}
\begin{eqnarray}
O_\kappa &=& \frac{1}{LS^2} \sum_l \kappa_l \ne 0, \label{eq:Ochl} \\
\kappa_l &=& S^x_l S^y_{l+1} - S^y_l S^x_{l+1} 
= \left[ {\mib S}_l \times {\mib S}_{l+1} \right]_z , \nonumber
\end{eqnarray}
where ${\mib S}_l$ is a spin-$S$ operator at site $l$ 
and $L$ is the system size assumed to be even throughout this paper.
The chiral phase breaks only the parity symmetry spontaneously 
with preserving the time-reversal and translational symmetries.
Using the bosonization method, Nersesyan {\it et al.} predicted 
the appearance of the chiral phase 
in the frustrated antiferromagnetic $S = 1/2$ chain 
with nearest-neighbor and next-nearest-neighbor couplings 
of easy-plane anisotropy.~\cite{Ner}
From the analytical~\cite{Leche,Kole} and 
numerical~\cite{KKH,HKK1,HKK2} studies,
it has been shown that the chiral phase appears 
in the frustrated chain with the anisotropic couplings 
in general spin-$S$ cases.
The ground-state phase diagrams for the $S = 1/2, 1, 3/2$, and $2$ cases 
have been determined numerically.~\cite{HKK1,HKK2}

Another interesting observation 
in the studies~\cite{Kole,KKH,HKK1,HKK2} is that 
{\it there exist two different chiral phases} 
in the integer-$S$ cases.
One of them is the ^^ ^^ chiral-Haldane" phase where 
the chiral LRO coexists with the string LRO~\cite{str,gstr}
\begin{equation}
O_{\rm str} = \frac{1}{LS} \sum_l \exp \left(
   \sum_{k=1}^{l-1} {\rm i} \frac{\pi}{S} S^z_k \right) S^z_l, 
\label{eq:Ostr}
\end{equation}
which characterizes the well-known Haldane phase,~\cite{Hal1,Hal2} 
and the spin correlation function decays exponentially 
with a finite energy gap.
The other is the ^^ ^^ gapless chiral" phase where 
the chiral LRO exists and the spin and string correlation functions 
decay algebraically with gapless excitations.
It has been shown that for $S = 1$ and $2$ 
the gapless chiral phase exists in a broad region of the ground-state 
phase diagrams while the chiral-Haldane phase exists 
in a narrow region between the Haldane and gapless chiral phases. 
(See Fig. 1 in ref. \citen{HKK1} and Fig. 9 in ref. \citen{HKK2}.)
For $S = 1/2$ and $3/2$, on the other hand, 
the gapless chiral phase exists as in the integer-$S$ cases 
while the chiral phase with a finite energy gap has not been identified.

As the mechanism generating the chiral phases,
easy-plane anisotropy plays an essential role as well as frustration does.
The anisotropy lowers the symmetry of the system 
from $SU(2)$ to $U(1) \times Z_2$ 
and induces a two-fold discrete degeneracy 
according as the right- and left-handed chirality.
In fact, the chiral phases do not appear in the isotropic case 
where the discrete chiral degeneracy no longer exists.
From the experimental viewpoint, 
the anisotropy in real materials commonly takes the form of 
the uniaxial single-ion-type.
It is therefore important to study the effect 
of the single-ion-type anisotropy on the chiral phases, 
for the sake of exploring experimental implications 
of the theoretical results and stimulating an investigation 
into the chiral phases in real materials.
Nevertheless, most of the theoretical findings on the chiral phases 
mentioned above 
have been obtained for the case of the anisotropic exchange couplings.
Kolezhuk~\cite{Kole} studied recently the frustrated 
integer-$S$ spin chain with the single-ion-type anisotropy 
by means of a large-$S$ approach 
and presented a schematic phase diagram 
including the gapless chiral phase and a chiral phase 
with a finite energy gap.
However, systematic studies on the chiral phases 
in the case of single-ion-type anisotropy have been scarce so far.

The aim of this paper is to investigate 
the ground-state properties of the frustrated $S = 1$ Heisenberg chain 
with the uniaxial single-ion-type anisotropy.
The model Hamiltonian is given by
\begin{equation}
{\cal H} = \sum_{\rho = 1,2} \left\{ J_\rho 
\sum_l {\mib S}_l \cdot {\mib S}_{l+\rho} \right\}
+ D \sum_l \left( S^z_l \right)^2.   \label{eq:Ham}
\end{equation}
I concentrate on the case of the antiferromagnetic couplings 
($J_1 > 0$ and $J_2 > 0$) and the easy-plane anisotropy ($D > 0$).
Hereafter I denote $j \equiv J_2/J_1$ and $d \equiv D/J_1$.
The ground-state properties of the model (\ref{eq:Ham}) 
have been studied for some limiting cases.
For $j = 0$ the phase transition between the Haldane and 
large-D (LD) phases have been studied 
intensively.~\cite{Hal-LD1,Hal-LD2,Hal-LD3,Hal-LD4}
It has been found that as $d$ increases the system undergoes 
a continuous phase transition from the Haldane phase to the LD phase
at $d = 1.000 \pm 0.001$.~\cite{Hal-LD4}
Meanwhile, it has been shown for $d = 0$ that 
as $j$ increases the system undergoes a first order transition 
at $j \simeq 0.747$ from the Haldane phase 
to the ^^ ^^ double Haldane" (DH) phase~\cite{HKK1,DH1,DH2}, in which 
the next-nearest-neighbor coupling $J_2$ becomes dominant so that 
the system is described as two Haldane subchains 
weakly coupled by the inter-subchain coupling $J_1$.
The string LRO vanishes discontinuously at the transition.
Furthermore, the appearance of the gapless and gapped chiral phases 
has been predicted by the large-$S$ approach.~\cite{Kole}

In the present work, I determine numerically the ground-state phase diagram 
of the model (\ref{eq:Ham}) by employing the same method as the one used 
in our previous works~\cite{KKH,HKK1,HKK2}: 
I calculate the appropriate correlation functions 
associated with the order parameters characterizing each phase 
using the density-matrix renormalization group (DMRG) 
method,~\cite{White1,White2} 
and then, analyze their long-distance behaviors.
From the obtained phase diagram, we find that 
the gapless chiral phase appears in a broad region of the $j$-$d$ plane 
while the chiral-Haldane phase appears 
in a narrow region between the Haldane and gapless chiral phases, 
which are essentially the same as those in the case 
of the anisotropic exchange couplings. 
It is found further that, in addition to the 
gapless chiral and chiral-Haldane phases, 
another novel chiral phase exists between 
the gapless chiral and LD phases.
This ^^ ^^ chiral-LD" phase 
is characterized by the chiral LRO and 
the exponential decay of both the spin and string correlation functions.

The paper is organized as follows.
In \S 2, I introduce the chiral, string, and spin correlation 
functions associated with each order parameter 
and explain the numerical method.
The obtained phase diagram and numerical data are presented in \S 3.
The results are summarized in \S 4.

\section{Correlation Functions and Numerical Method}
The method used to determine the phase diagram is 
essentially the same as those used in our previous works.~\cite{KKH,HKK1,HKK2}
I have calculated the two-point chiral, string, 
and spin correlation functions defined by 
\begin{eqnarray}
C_\kappa(l,l') &=& \langle \kappa_l \kappa_{l'} \rangle, \label{eq:Cchl} \\
C_{\rm str}(l,l') &=& \langle S^z_l \exp \left(
{\rm i}\pi \sum_{k=l}^{l'-1}S^z_k \right) S^z_{l'} \rangle, \label{eq:Cstr} \\
C_s^x(l,l') &=& \langle S^x_l S^x_{l'} \rangle, \label{eq:Cxx}
\end{eqnarray}
which are associated with the order parameters (\ref{eq:Ochl}), 
(\ref{eq:Ostr}), and (\ref{eq:Ohel}), respectively.
The notation $\langle \cdots \rangle$ represents 
the expectation value in the lowest energy state 
in the subspace of $S^z_{\rm total} = \sum_l S^z_l = 0$ with even parity.
It has been checked by the exact-diagonalization calculation 
that the ground state of the chain indeed belongs to the subspace.
The calculation has been performed for various fixed values of $d$ ($j$) 
with varying $j$ ($d$).
Then, I have estimated the transition points $j_c$ ($d_c$) 
by examining the dependence of the correlation functions on $r = |l-l'|$ 
at long distance.

I have employed the infinite-system DMRG algorithm 
originally introduced by White~\cite{White1,White2} 
and accelerated it by making use of the recursion relation 
proposed by Nishino and Okunishi.~\cite{PWFRG1,PWFRG2}
The number of kept states $m$ is up to $350$ if not otherwise mentioned.
The convergence of the data with respect to $m$ has been checked 
by increasing $m$ consecutively.
Since the truncation error of the DMRG calculation 
increases dramatically as $j$ increases, 
the calculation is limited to rather small $j$.
However, I consider that the region of $j$ treated in this work, $j \lsim 1$, 
is enough to study the properties of the chiral phases 
which we wish to know.
In the calculations, the open boundary condition has been imposed
for the sake of the best performance of the algorithm.
Accordingly, in order to avoid the unwanted boundary effect, 
the two-point correlation functions have been calculated 
for the sites $l$ and $l'$ near the center of the chain, i.e., 
$l = l_0 - r/2$ and $l' = l_0 + r/2$ where 
$l_0 = L/2$ for even $r$ and $l_0 = (L+1)/2$ for odd $r$.
By checking the convergence of the data with increasing $L$ 
typically up to 2000, 
I have confirmed that the data are free from the effect of 
the open boundaries.

\section{Phase Diagram}
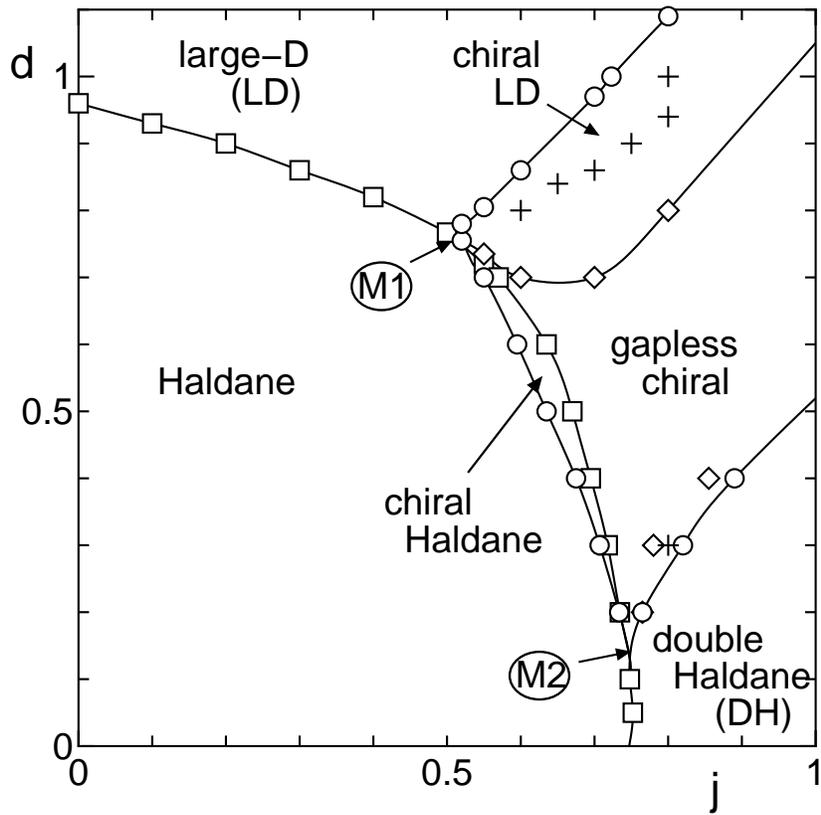
\begin{figure}
\caption{The ground-state phase diagram of the model (\ref{eq:Ham}).
The circles, squares, and diamonds represent the points where 
the chiral LRO sets in, the points where the string LRO vanishes, 
and the points where the behavior of the spin and string correlations 
changes from an exponential to an algebraic decay.
The multi-critical points $(j_{M1},d_{M1})$ and $(j_{M2},d_{M2})$ 
are indicated as ^^ ^^ M1" and ^^ ^^ M2", respectively.
The crosses represent the points for which the calculations with $m = 500$ 
are performed.
The lines are to guide the eye.
}
\label{fig:diag}
\end{figure}

By examining the long-distance behaviors of 
the chiral, string, and spin correlation functions 
introduced in the previous section,
I have determined the ground-state phase diagram 
of the model (\ref{eq:Ham}) in the $j$-$d$ plane.
The obtained phase diagram shown in Fig. \ref{fig:diag} 
includes six different phases, i.e., 
the Haldane, LD, DH, gapless chiral, chiral-Haldane, and chiral-LD phases.
The long-distance behaviors of the correlation functions 
in each of the phases are summarized in Table \ref{tab:cor}.
We can see in Fig.\ref{fig:diag} 
that the gapless chiral phase appears in a broad region 
of the phase diagram while the chiral-Haldane phase appears 
between the Haldane and gapless chiral phases.
Further we find the chiral-LD phase, where the chiral LRO exists 
while the spin and string correlation decay exponentially, 
between the LD and gapless chiral phases.
Two multi-critical points, at least, are observed: 
One of them is at $(j_{M1},d_{M1}) \simeq (0.51,0.76)$ 
among the Haldane, LD, gapless chiral, chiral-Haldane, and chiral-LD phases
while the other is at $(j_{M2},d_{M2}) \simeq (0.75,0.15)$ 
among the Haldane, gapless chiral, chiral-Haldane and DH phases.
The numerical data for each transition line 
are discussed in the following subsections.

\subsection{Haldane-chiral transition}
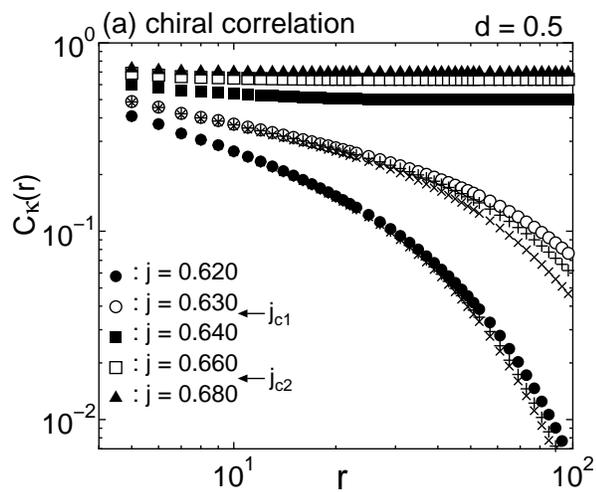
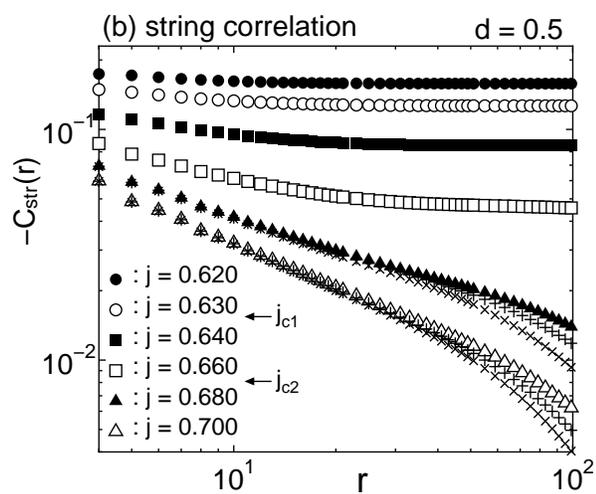
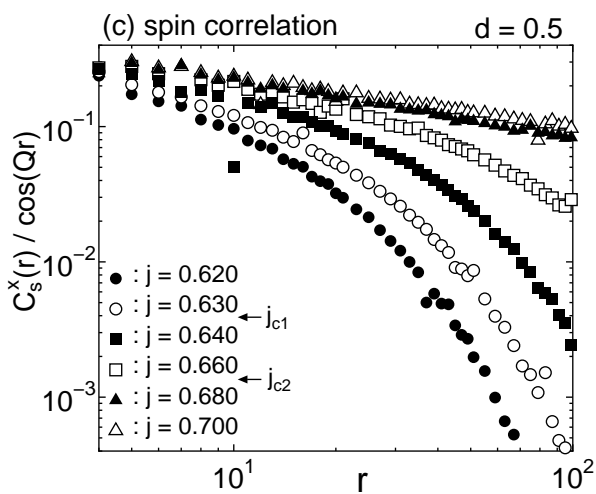
\begin{figure}
\caption{The $r$ dependence of the correlation functions for $d = 0.5$: 
(a) chiral correlation $C_\kappa(r)$; 
(b) string correlation $-C_{\rm str} (r)$;
(c) spin correlation $C_s^x(r)$ divided by the oscillating factor $\cos(Qr)$.
The number of kept states is $m=350$.
To illustrate the $m$ dependence, the data with $m= 260$ and $300$ are 
indicated by crosses: for $j = 0.620$ and $0.630$ in (a); 
for $j = 0.680$ and $0.700$ in (b).
In other cases, the truncation errors are smaller than the symbols.
The data of $C_\kappa (r)$ for $j = 0.700$ in (a), which are almost 
the same as those for $j = 0.680$, are omitted for clarity.}
  \label{fig:H-C}
\end{figure}

First let us consider the phase transitions between the Haldane 
and chiral phases.
Figures \ref{fig:H-C}(a)-(c) show the data of the chiral, string, 
and spin correlation functions in log-log plots 
for $d = 0.5$ and several typical values of $j$.
The data of the spin correlation shown in Fig. \ref{fig:H-C} (c) 
are divided by the leading oscillating factor $\cos(Qr)$ 
where $Q$ is the wavenumber characterizing 
the incommensurate oscillation of $C_s^x(r)$ in real space.
It can be clearly seen in Fig. \ref{fig:H-C} (a) that 
the data of the chiral correlation function $C_\kappa(r)$ 
are bent upward for $j > j_{c1} \simeq 0.635$ 
suggesting a finite chiral LRO 
while they are bent downward for $j < j_{c1}$ 
suggesting the absence of the chiral ordering.
The transition point where the chiral LRO sets in 
is estimated to be $j_{c1} = 0.635 \pm 0.005$.
As shown in Fig. \ref{fig:H-C} (b), the string correlation function 
exhibits a finite LRO for $j < j_{c2} \simeq 0.670$ 
whereas it decays algebraically for $j > j_{c2}$.~\cite{trunc}
Meanwhile, the data of the spin correlation function 
shown in Fig. \ref{fig:H-C} (c) 
are bent downward for $j < j_{c2}$ suggesting an exponential decay 
with a finite energy gap 
while they exhibit a linear behavior for $j > j_{c2}$ 
suggesting a power-law decay with gapless excitations.
The transition point $j_{c2}$ where the string LRO and the energy gap 
vanish is estimated to be $j_{c2} = 0.670 \pm 0.010$.
It should be noticed that the estimate of $j_{c2}$ is 
distinctly larger than that of $j_{c1}$. 
Accordingly, the chiral-Haldane phase 
where the chiral and string LROs coexist and 
the spin correlation decays exponentially exists 
in a narrow but finite region between the Haldane and gapless chiral phases.
The existence of the chiral-Haldane phase can be clearly seen 
from the behavior of the correlation functions at $j = 0.640$ and $0.660$
which lie between $j_{c1}$ and $j_{c2}$.
Thus, it is concluded that the system undergoes two successive transitions 
as $j$ increases, first at $j = j_{c1}$ from the Haldane phase to 
the chiral-Haldane phase, and then, at $j = j_{c2}$ 
to the gapless chiral phase.
The results are essentially the same as the ones observed 
in the frustrated chain with anisotropic exchange couplings.~\cite{KKH,HKK1}

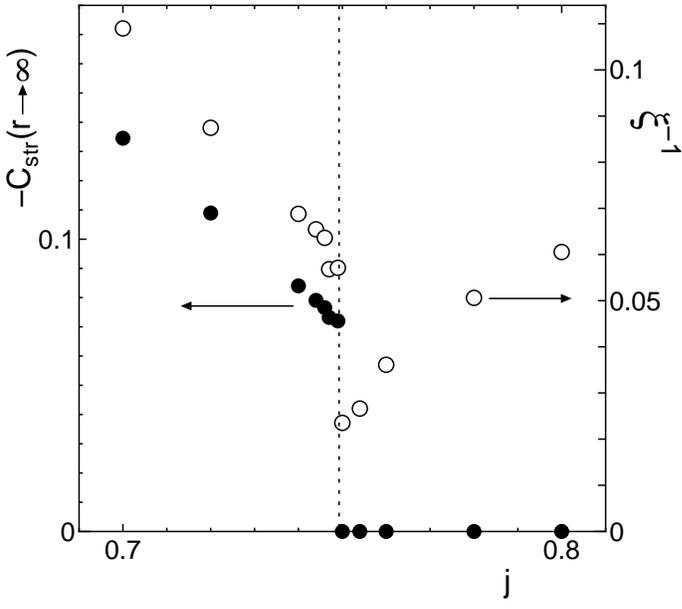
\begin{figure}
\caption{The $j$ dependence of the string correlation 
function $-C_{\rm str}(r)$ at $r \to \infty$ (solid circle) and
the inverse spin-correlation length $\xi^{-1}$ (open circle) for $d = 0.1$.
The dotted line represents the transition point.}
  \label{fig:order}
\end{figure}

Performing calculations in the same manner for various fixed $d$ ($j$)
with varying $j$ ($d$), I have determined the transition points 
$j_{c1}$ ($d_{c1}$) and $j_{c2}$ ($d_{c2}$) plotted in Fig. \ref{fig:diag}.
The figure may suggest that the transition lines connecting 
the estimated points run from the multi-critical point $(j_{M1},d_{M1})$ 
to the other multi-critical point $(j_{M2},d_{M2})$.
The chiral-Haldane phase is identified for 
$d = 0.7, 0.6, 0.5, 0.4$, and $0.3$ 
while it is not for $d = 0.2$ within the calculation 
with varying $j$ at intervals of $0.002$.
However, I note that it is difficult to exclude the possibility that 
the chiral-Haldane phase exists in a too narrow region to be detected.
Thus it is not clear whether the transition lines $j_{c1}$ and $j_{c2}$ merge 
at $(j_{M2},d_{M2})$ or at another multi-critical point $(j'_{M2},d'_{M2})$ 
with $d'_{M2} > 0.2$.
In the latter case, a transition line $j'_{c1}$ 
between the Haldane and gapless chiral phases connects $(j'_{M2},d'_{M2})$ 
and $(j_{M2},d_{M2})$.
For $d = 0.1$ and $0.05$, on the other hand, 
the chiral LRO is not observed.
In these cases, it is also found that 
the string LRO vanishes discontinuously at the transition point and 
the spin correlation function exhibits an exponential decay 
for all calculated values of $j$.
The $j$ dependence of the string correlation function 
at $r \to \infty$ for $d = 0.1$ is shown in Fig. \ref{fig:order} 
together with the spin-correlation length $\xi$ estimated by fitting 
the data of $C_s^x(r)$ to the form of an exponential decay, 
$C_s^x(r) = A \cos(Qr) \exp(-r/\xi)$, where $A$ is a numerical constant.
These behaviors of the string and spin correlation functions 
for $d = 0.1$ and $0.05$ suggest a first-order transition 
between two gapful phases, i.e., the Haldane and DH phases.
From the observations, it is concluded that between $d = 0.2$ and $d = 0.1$
there is a multi-critical point $(j_{M2},d_{M2})$ where 
the transition lines $j_{c1}$ and $j_{c2}$ 
or the transition line $j'_{c1}$ merges with 
the DH-chiral transition line discussed in \S 3.4.
The first-order transition line between the Haldane and DH phases 
smoothly connects the multi-critical point $(j_{M2},d_{M2})$ 
to the transition point in the isotropic case $(d = 0)$, 
$j_{c2} \simeq 0.747$, estimated in the previous works.~\cite{HKK1,DH1,DH2}.

\subsection{LD-chiral transition}
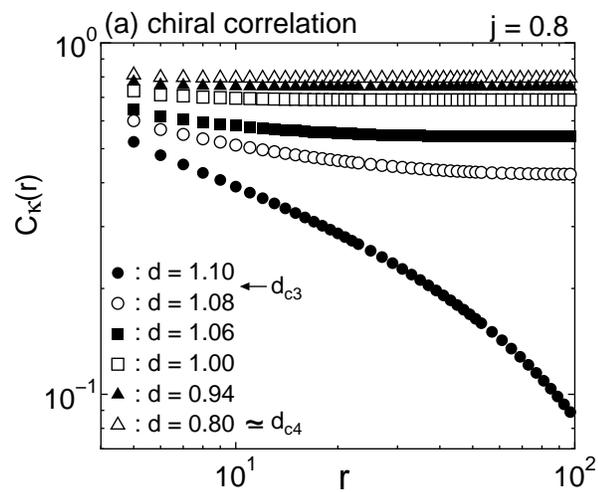
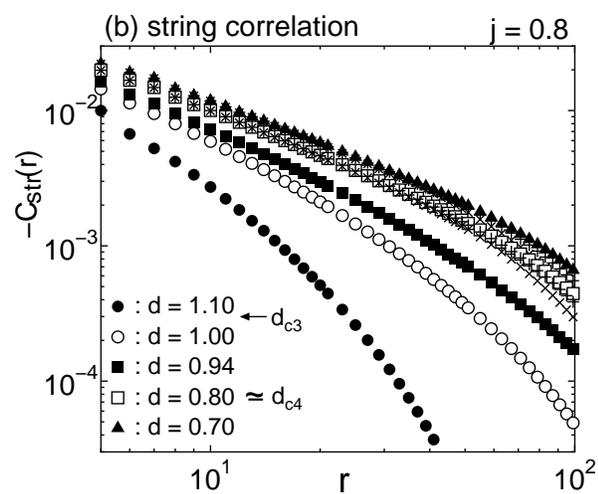
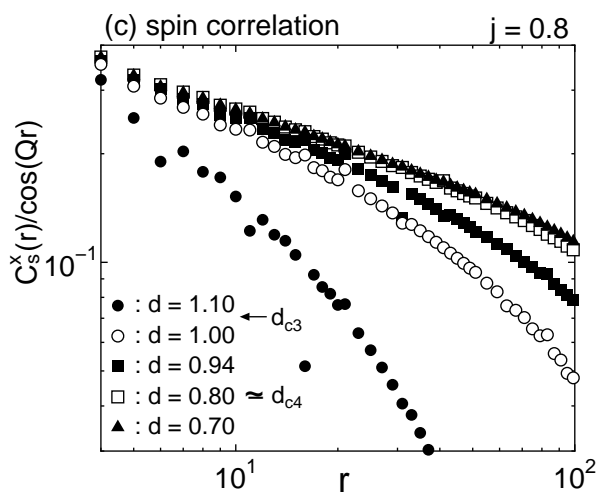
\begin{figure}
\caption{The $r$ dependence of the correlation functions for $j = 0.8$: 
(a) chiral correlation $C_\kappa(r)$; 
(b) string correlation $-C_{\rm str} (r)$;
(c) spin correlation $C_s^x(r)$ divided by the oscillating factor $\cos(Qr)$.
The number of kept states is $m=500$ for $d = 0.94$ and $1.00$ 
while it is $m=350$ for the other cases.
To illustrate the $m$ dependence, the data with $m=260$ and $300$ 
for $d = 0.80$, and $0.70$ are indicated in (b) by crosses.
In other cases, the truncation errors are smaller than the symbols.}
  \label{fig:LD-C}
\end{figure}

We next consider the phase transitions between the LD and chiral phases.
The data of the chiral, string, and spin correlation functions 
for $j = 0.8$ and several typical values of $d$ 
are shown in Figs. \ref{fig:LD-C} (a)-(c) in log-log plots.
As shown in Fig. \ref{fig:LD-C} (a), the chiral correlation function 
$C_\kappa(r)$ exhibits a finite LRO for $d < d_{c3} \simeq 1.09$ 
while it decays exponentially for $d > d_{c3}$.
The transition point $d_{c3}$ where the chiral LRO vanishes is 
thereby estimated as $d_{c3} = 1.09 \pm 0.01$.
Meanwhile, the string and spin correlations decay exponentially 
for $d > d_{c4} \simeq 0.80$ suggesting a finite energy gap 
whereas they decay algebraically for $d < d_{c4}$
suggesting gapless excitations.~\cite{trunc}
The transition point $d_{c4}$ is estimated to be $d_{c4} = 0.80 \pm 0.10$.
Unfortunately, the numerical error of the estimate of $d_{c4}$ is quite large 
due to the truncation error of the DMRG calculation, 
which tends to underestimate the string and spin correlation functions.
Nevertheless, from the behaviors of the correlation functions 
for $0.94 \lsim d < d_{c3}$, it can be clearly seen that 
there exists a intermediate region 
with a finite chiral LRO and a finite energy gap.
In order to confirm the appearance of the intermediate phase, 
I have performed a high-precision calculation with $m$ up to $500$ 
for $d = 1.00$ and $0.94$.
The data shown in Fig. \ref{fig:LD-C}, 
which are almost free from the truncation error, 
clearly suggest that the parameter points $d = 1.00$ and $0.94$ 
indeed belong to the chiral phase with a finite energy gap.
It is therefore concluded that the estimate of $d_{c4}$ is distinctly smaller 
than that of $d_{c3}$ and the system undergoes two successive transitions 
as $d$ decreases, first at $d = d_{c3}$ from the LD phase to 
the ^^ ^^ chiral-LD" phase characterized by a finite chiral LRO 
and an exponential decay of the string and spin correlation functions,
and then at $d = d_{c4}$ to the gapless chiral phase.
Here it should be noticed that the chiral-LD phase is distinct 
from the chiral-Haldane phase with respect to the absence of the string LRO.
Hence the chiral-LD phase can be regarded as a new type of 
the gapped chiral phase.

By the same calculation for various fixed $j$ ($d$), the transition points 
$d_{c3}$ ($j_{c3}$) and $d_{c4}$ ($j_{c4}$) are estimated.
In particular, the appearance of the chiral-LD phase has been confirmed 
by the high-precision calculations with $m$ up to $500$ 
for the points plotted in Fig. \ref{fig:diag} by crosses.
As seen in Fig. \ref{fig:diag}, the chiral-LD phase 
appears in a finite region between the LD and gapless chiral phases.
The transition lines $d_{c3}$ and $d_{c4}$ 
merge with the Haldane-chiral transition lines $j_{c1}$ and $j_{c2}$ 
around the multi-critical point $(j_{M1},d_{M1})$
into the Haldane-LD transition line discussed in the next subsection.

\subsection{Haldane-LD transition}
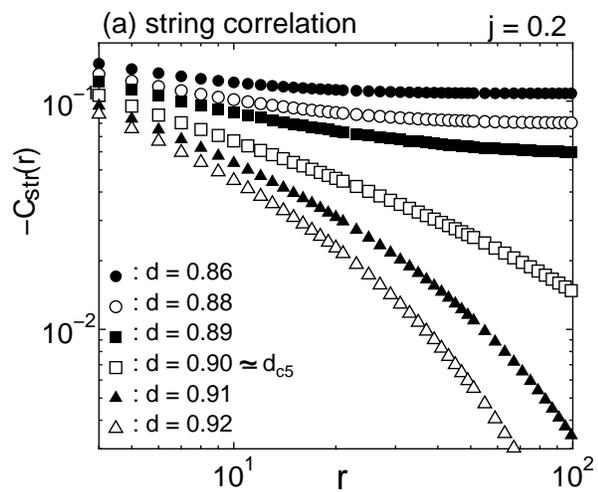
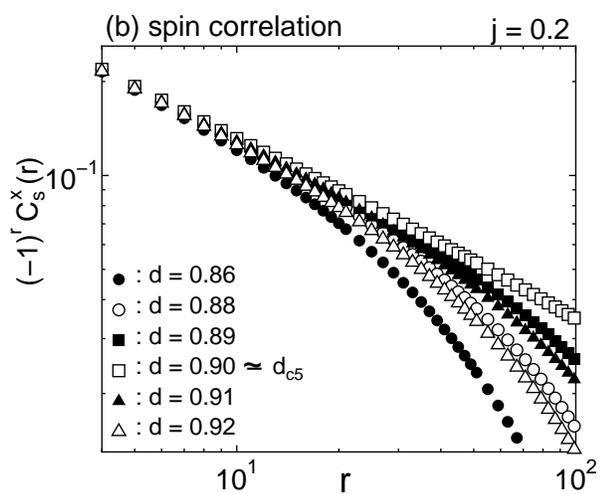
\begin{figure}
\caption{The $r$ dependence of the correlation functions for $j = 0.2$: 
(a) string correlation $-C_{\rm str} (r)$;
(b) spin correlation $(-1)^r C_s^x(r)$.
The number of kept states is $m=180$.
The truncation errors are smaller than the symbols.}
\label{fig:H-LD} 
\end{figure}

In this subsection, we see the results for the phase transition 
between the Haldane and LD phases.
The phase transition has been studied in detail 
in the case of no frustration ($j = 0$).~\cite{Hal-LD1,Hal-LD2,Hal-LD3,Hal-LD4}
It has been pointed out that there occurs a continuous phase transition 
accompanying the vanishing of the energy gap
at $d = 1.000 \pm 0.001$.
In Figs. \ref{fig:H-LD} (a) and (b), I show the data of 
the string and spin correlation functions for $j = 0.2$ and 
several typical values of $d$.
Here the $m$ convergence of the data has almost been achieved 
even at $m = 180$.
We can see in Fig. \ref{fig:H-LD} (a) that 
the string correlation function 
exhibits a finite LRO for $d < d_{c5} \simeq 0.90$ 
while it exhibits an exponential decay for $d > d_{c5}$.
In the meantime, the spin correlation function exhibits 
an exponential decay for $d \le 0.89$ and $d \ge 0.91$
in accordance with the existence of a finite energy gap in both phases 
while it decays algebraically at $d = 0.90$ 
suggesting the vanishing of the energy gap at the transition.
The transition point between the Haldane and LD phases 
for $j = 0.2$ is thereby estimated to be $d_{c5} = 0.90 \pm 0.01$.
The transition points $d_{c5}$ estimated in the same way 
are plotted in Fig. \ref{fig:diag}.
The transition line connecting the estimates 
runs from the point $d_{c5} = 0.96 \pm 0.02$ at $j = 0$, 
which is slightly smaller than the previous estimate 
$d_{c5}(j=0) = 1.000 \pm 0.001$,
to the multi-critical point $(j_{M1},d_{M1})$.

\subsection{DH-chiral transition}
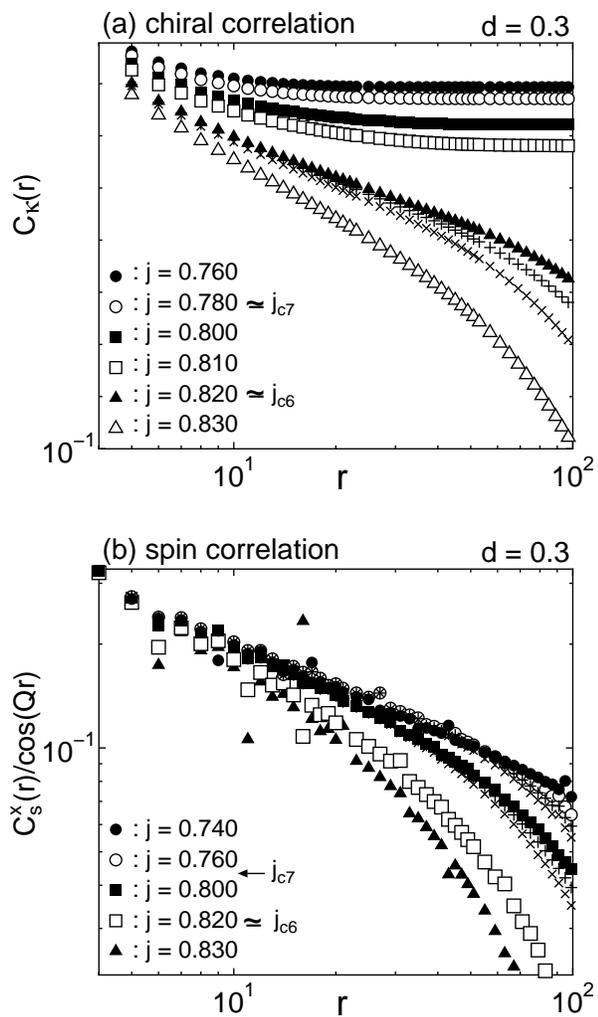
\begin{figure}
\caption{The $r$ dependence of the correlation functions for $d = 0.3$: 
(a) chiral correlation $C_\kappa(r)$; 
(b) spin correlation $C_s^x(r)$ divided by the oscillating factor $\cos(Qr)$.
The number of kept states is $m = 500$ for $j = 0.800$ and $m = 350$
for other values of $j$.
To illustrate the $m$ dependence, the data with smaller $m$ are 
indicated by crosses: the data with $m= 260$ and $300$ for $j=0.820$ in (a); 
the data with $m = 350$ and $400$ for $j = 0.800$ and 
with $m=260$ and $300$ for $j = 0.760$ in (b).
In other cases, the truncation errors are smaller than the symbols.
}
\label{fig:DH-C}
\end{figure}

Here we consider the phase transition between the DH and chiral phases.
Figure \ref{fig:DH-C} (a) shows the chiral correlation 
function $C_\kappa (r)$ for $d = 0.3$ and several typical values of $j$ 
around the DH-chiral transition.
As shown in the figure, $C_\kappa (r)$ exhibits a finite LRO 
for $j < j_{c6} \simeq 0.820$ 
while it decays exponentially for $j > j_{c6}$.
Thus, the transition point where the chiral LRO vanishes is 
estimated to be $j_{c6} = 0.820 \pm 0.010$.
Meanwhile, the data of the spin correlation function for $d = 0.3$ are 
shown in Fig. \ref{fig:DH-C} (b) where the data are divided 
by the oscillating factor $\cos(Qr)$.
It can be seen in the figure that as $j$ increases the spin correlation 
changes its behavior from an algebraic decay to 
an exponential decay at $j = j_{c7} \simeq 0.780$.
Although the truncation error of the data prevents us from 
carrying out the precise estimation, 
the transition point is estimated to be $j_{c7} = 0.780 \pm 0.040$.

Here arises a question: Whether $j_{c7}$ is equal to 
or smaller than $j_{c6}$ ?
If $j_{c7}$ is equal to $j_{c6}$, there occurs only one phase transition 
at $j = j_{c6} = j_{c7}$ between the gapless chiral phase and the DH phase.
On the other hand, if $j_{c7}$ is smaller than $j_{c6}$, 
the system undergoes two successive transitions as $j$ increases, 
first at $j = j_{c6}$ from the gapless chiral phase 
to an intermediate ^^ ^^ chiral-DH" phase 
with a finite chiral LRO and a finite energy gap, 
and then at $j = j_{c7}$ to the DH phase.
Unfortunately, the estimates of the transition points,
particularly that of $j_{c7}$, are not 
accurate enough to determine which of the cases is realized. 
Here I note that the data of the spin correlation 
for $j = 0.800$ shown in Fig. \ref{fig:DH-C} (b) 
are calculated with $m$ up to $500$ 
and still suffer from the truncation error which is not negligible.
This means that it is unpromising to achieve 
the sufficiently accurate estimate of $j_{c7}$ 
in the manner used in this work.
The difficulty in identifying the intermediate phase 
in this case can be ascribed partly to the fact that no order parameter 
characterizing the DH phase has been found so far~\cite{DHorder}
and partly to the fact that the intermediate phase, 
if any, appears only in a narrow region 
which can be concealed behind the error bar of $j_{c7}$.
Hence, the situation here is different from 
that of the Haldane-chiral transition where we can utilize 
the string order parameter to estimate $j_{c2}$ 
and that of the LD-chiral transition where the region of 
the chiral-LD phase is so broad that we can identify the chiral-LD phase 
even with rather rough estimates of $d_{c4}$.
Since it is a subtle problem to estimate the point 
where the behavior of the spin correlation changes, 
it might be needed for the settlement of the question 
to introduce an order parameter characterizing the DH phase.
This problem remains open for future studies.

In the same manner, I have estimated the transition points 
$j_{c6}$ and $j_{c7}$ for $d = 0.2$ and $d = 0.4$.
The chiral-DH phase has not been identified also in these cases.
As shown in Fig. \ref{fig:diag}, the transition lines connecting 
the estimates seem to emerge from the multi-critical point $(j_{M2},d_{M2})$.
The values of $j_{c6}$ and $j_{c7}$ become larger as $d$ becomes larger 
suggesting that the DH phase becomes stable as $j$ increases.

\subsection{Multi-critical point}
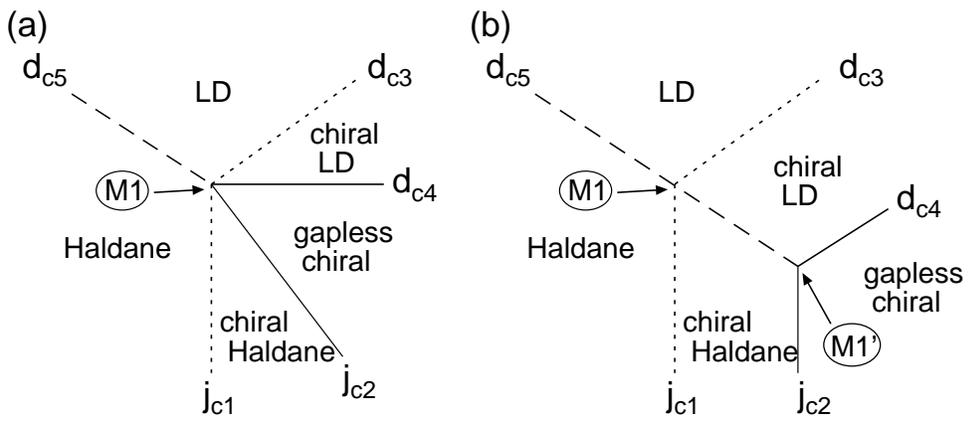
\begin{figure}
\caption{The possible schematic pictures of the phase diagram 
around the multi-critical point $(j_{M1},d_{M1})$.
Figures (a) and (b) correspond to the cases (i) and (ii) in text, 
respectively.
}
\label{fig:multi}
\end{figure}

Lastly, I touch upon the region of the phase diagram
around the multi-critical point $(j_{M1},d_{M1})$.
It can be seen in the obtained phase diagram Fig. \ref{fig:diag} 
that the five transition lines run into a narrow region 
around $(j_{M1},d_{M1})$.
It might be useful to classify them into three groups, i.e., 
the transition lines $j_{c1}$ and $d_{c3}$ where the chiral LRO sets in, 
the lines $j_{c2}$ and $d_{c4}$ between gapful and gapless phases, 
and the line $d_{c5}$ between the Haldane and LD phases.
Here I note that the mechanism generating the chiral LRO 
is essentially distinct from 
the one generating the energy gap~\cite{Leche,Kole} 
and, to my knowledge, no definite relation between them has been known.
Thus, the behavior of the chiral transition lines ($j_{c1}$ and $d_{c3}$) 
and the ^^ ^^ gapful-gapless" lines ($j_{c2}$ and $d_{c4}$) around 
the multi-critical point is still nontrivial.
Looking at Fig. \ref{fig:diag}, 
two possibilities seem to be left: 
(i) The five transition lines merge at 
one multi-critical point $(j_{M1},d_{M1})$.
(ii) The transition lines $j_{c1}$ and $d_{c3}$
merge with the Haldane-LD transition line $d_{c5}$ at $(j_{M1},d_{M1})$ 
while the lines $j_{c2}$ and $d_{c4}$ 
merge with the transition line $d'_{c5}$ between 
the chiral-Haldane and chiral-LD phases 
at another multi-critical point $(j'_{M1},d'_{M1})$.
These scenarios are illustrated in Fig. \ref{fig:multi} (a) and (b), 
respectively.
Because of the rather large error of 
the estimates of $d_{c4}$ mentioned in \S 3.2,
we can not decide which of the cases is realized 
from the present numerical results.
More detailed studies might be needed to identify the universality class 
of the multi-critical point(s) and transition lines.

\section{Summary}
In this paper, the ground-state properties of 
the frustrated $S=1$ Heisenberg spin chain with 
uniaxial single-ion-type anisotropy (eq. (\ref{eq:Ham})) 
have been investigated numerically by the infinite-system DMRG method.
By calculating the chiral, string, and spin correlation functions 
and analyzing their long-distance behaviors, 
I have determined the ground-state phase diagram for $0 \le j \lsim 1$ 
and $d \ge 0$ (Fig. \ref{fig:diag}).
It has turned out that there exist six different phases, 
namely, the Haldane, LD, DH, gapless chiral, 
chiral-Haldane, and chiral-LD phases.
The gapless chiral phase appears in a broad region of the phase diagram 
while the chiral-Haldane phase appears in the narrow but finite region 
between the Haldane and gapless chiral phases; 
these results are essentially the same as those found in the frustrated 
$S = 1$ chain with anisotropic exchange couplings.~\cite{KKH,HKK1} 
Meanwhile, the chiral-LD phase has been found 
between the LD and gapless chiral phases.
Since the chiral-LD phase is characterized by 
a finite chiral LRO, the absence of the string LRO, and 
the short-range spin correlation function,
it is distinct from either the gapless chiral or the chiral-Haldane phase 
and can be regarded as a new type of the gapped chiral phases.
The question about the existence of the possible chiral-DH phase 
remains open.

After the theoretical finding of the chiral phases, 
one of the most challenging tasks is the experimental observation of  
the phases in real materials.
As seen in the obtained phase diagram, 
the anisotropy $d \gsim 0.15$ and the next-nearest neighbor coupling $j$ 
within a suitable range around $j \simeq 0.75$ 
are required for a material realizing the chiral phases.
These values of $d$ and $j$ are realistic so that 
there must be a good chance to find out a material in the parameter region.
If such a material is prepared, 
the measurement of the vector chirality (\ref{eq:Ochl}) 
is, in principle, possible by using polarized neutrons.~\cite{pn1,pn2,pn3}
I hope that the results of the present paper stimulate 
further experimental studies.

\section*{Acknowledgments}
The author is very grateful to M. Kaburagi and H. Kawamura 
for collaboration in early stage of this work.
He also thanks X. Hu for useful discussions.
Numerical calculations were carried out in part 
at Yukawa Institute Computer Facility, Kyoto University.
The author was supported by a Grant-in-Aid for Encouragement 
of Young Scientists from Ministry of Education, Science and Culture of Japan.

\begin{table}
\caption{The long-distance behaviors of the spin, chiral, and string 
correlation functions in each of the phases.
The words "expo." and "power" mean an exponential decay 
and a power-law decay, respectively.} 
\label{tab:cor}
\begin{tabular}{lcccccc}
             &Haldane  &chiral-Haldane&gapless chiral& LD  &chiral-LD & DH \\
\hline
$C_{\rm s}^x$&  expo.  &   expo.     &     power    &expo.&  expo.  & expo.\\
$C_\kappa$   &  expo.  &   LRO       &     LRO      &expo.&  LRO    & expo.\\
$C_{\rm str}$&  LRO    &   LRO       &     power    &expo.&  expo.  & expo.\\
\end{tabular}
\end{table}%


\begin{thebibliography}{99}
\bibitem{scalchi}
Note that the vector chirality (\ref{eq:Ochl}) differs from 
the scalar chirality 
$\chi_l = {\mib S}_{l-1} \cdot {\mib S}_l \times {\mib S}_{l+1}$ 
which is often discussed in the context of isotropic spin systems: 
The vector chirality changes its sign under the parity operation 
but is invariant under the time-reversal operation 
whereas the scalar chirality changes its sign under both operations.
\bibitem{trunc}
The string correlation functions for $j = 0.680$ and $0.700$ 
in Fig. \ref{fig:H-C} (b) 
and for $d = 0.80$ and $0.70$ in Fig. \ref{fig:LD-C} (b) 
seem to be bent downward for $r \gsim 40$.
However, I note that the $m$ convergence of the data for $r \gsim 40$ 
has not been achieved yet 
and the string correlation is enhanced as $m$ increases.
Thus, from the behavior of the string correlation for $r \lsim 40$ 
where the $m$ convergence is almost achieved, 
I consider that the string correlations for the above $j$'s and $d$'s exhibit 
an algebraic decay and the bend-downward behavior observed 
for larger $r$ is an artifact due to the truncation error 
eventually vanishing in the limit $m \to \infty$.
\bibitem{DHorder}
Since the system in the DH phase can be described as two Haldane subchains, 
one might expect that the string correlation function 
on each subchains can be the order parameter of the phase.
Contrary to the naive expectation, however, 
the string correlation on a subchain does not exhibit a finite LRO 
but decays exponentially. (See refs.\citen{DH1} and \citen{DH2}.)
\end{thebibliography}
\end{document}